\documentclass[11pt]{article}
\usepackage{amsmath,amssymb,amsfonts,bbm}

\newcommand{\be}{\begin{equation}}
\newcommand{\ee}{\end{equation}}
\newcommand{\bea}{\begin{eqnarray}}
\newcommand{\eea}{\end{eqnarray}}
\newcommand{\ba}{\begin{array}}
\newcommand{\ea}{\end{array}}

\newcommand{\htwo}{h_{2,1}}
\newcommand{\M}{\mathcal{M}}
\newcommand{\N}{\mathcal{N}}
\newcommand{\D}{\mathcal{D}}
\newcommand{\K}{\mathcal{K}}
\newcommand{\F}{\mathcal{F}}
\newcommand{\A}{\mathcal{A}}

\long\def\symbolfootnote[#1]#2{\begingroup%
\def\thefootnote{\fnsymbol{footnote}}\footnote[#1]{#2}\endgroup}

\begin{document}

\thispagestyle{empty}\vspace{40pt}

\hfill{}

\vspace{128pt}

\begin{center}
    \textbf{\Large Symplectic covariance of the $\N=2$ hypermultiplets}\\
    \vspace{40pt}

    Moataz H. Emam\symbolfootnote[1]{\tt moataz.emam@cortland.edu}

    \vspace{12pt}   \textit{Department of Physics}\\
                    \textit{SUNY College at Cortland}\\
                    \textit{Cortland, NY 13045, USA}\\
\end{center}

\vspace{40pt}

\begin{abstract}

The main objective of this article is to recast the
hypermultiplets sector of five dimensional ungauged $\N=2$
supergravity into a manifestly symplectic-covariant form. We
propose that this facilitates the construction and analysis of
hypermultiplet fields coupled to $p$-brane sources and discuss examples.

\end{abstract}

\newpage


\vspace{15pt}

\section{Introduction}

The study of $\N=2$ supergravity (SUGRA) theories has gained
interest in recent years for a variety of reasons. For example,
$\N=2$ branes are particularly relevant to the conjectured
equivalence between string theory on anti-de Sitter space and
certain superconformal gauge theories living on the boundary of
the space (the AdS/CFT duality) \cite{Maldacena:2001uc}. Also
interesting is that many results were found to involve the
so-called attractor mechanism (\emph{e.g.}
\cite{hep-th/9508072,hep-th/9602111,hep-th/9602136}); the study of
which developed very rapidly with many intriguing outcomes
(\emph{e.g.} \cite{hep-th/0506177,hep-th/0507096,hep-th/0508042}).
The subject is also important in the context of string theory
compactifications, as it is known that the behavior of the lower
dimensional fields is contingent upon the topology of the
underlying submanifold. In addition, many $D=4,5$ results were
shown to be related to higher dimensional ones via wrapping over
specific cycles of manifolds with special holonomy. For example,
M-branes wrapping K\"{a}hler calibrated cycles of a Calabi-Yau
(CY) 3-fold \cite{Cho:2000hg} dimensionally reduce to black holes
and strings coupled to the vector multiplets of five dimensional
$\N=2$ supergravity \cite{Kastor:2003jy}, while M-branes wrapping
special Lagrangian calibrated cycles reduce to configurations
carrying charge under the hypermultiplet scalars
\cite{Martelli:2003ki,Fayyazuddin:2005as,Emam:2005bh,Emam:2006sr,Emam:2007qa}.
Studying how higher dimensional results are related to lower
dimensional ones may eventually provide clues to the explicit
structure of the compact space and the choice of compactification
mechanism, thereby contributing to more understanding of the
string theory landscape. It becomes then an important issue
indeed, as far as the string theoretic view of the universe is
concerned, to study such compactifications by classifying lower
dimensional solutions and analyzing how they relate to higher
dimensional ones.

In reviewing the literature, one notices that most studies in
$\N=2$ SUGRA in any number of dimensions specifically address the
vector multiplets sector; setting the hypermultiplets to zero.
This is largely due to the fact that the standard representation
of the hypermultiplet scalars as coordinates on a quaternionic
manifold is somewhat hard to deal with. It has been shown,
however, that certain duality maps relate the target space of a
given higher dimensional fields' sector to that of a lower
dimensional one \cite{Ferrara:1989ik}. Particularly relevant to
this work is the so-called c-map which relates the quaternionic
structure of the $D=5$ hypermultiplets to the more well-understood
special geometric structure of the $D=4$ vector multiplets. This
means that one can recast the $D=5$ hypermultiplet fields into a
form that makes full use of the methods of special geometry. This
was done in \cite{Gutperle:2000ve} and applied in the
same reference as well as in \cite{Emam:2005bh} and others. Using
this method, finding solutions representing the five dimensional
hypermultiplet fields often means coming up with ans\"{a}tze that
have special geometric form. This can be, and has been, done by
building on the considerable $D=4$ vector multiplets literature,
and in most cases the solutions are remarkably similar. For
example, $D=5$ hypermultiplet couplings to 2-branes and instantons \cite{Emam:2005bh, Gutperle:2000ve} lead to the same
type of attractor equations found for the vector multiplets
coupled to $D=4$ black holes (\emph{e.g.}
\cite{Behrndt:1997fq,Sabra:1997dh,Behrndt:1997ny,Behrndt:1998eq}).

Despite the power of the c-map method, it is still a highly
tedious process to find solutions representing the full set of
hypermultiplet fields. This is particularly serious in view of the
fact that the most general solutions necessarily depend on the
structure of the underlying Calabi-Yau manifold. Since no explicit
(nontrivial) compact CY 3-folds are known, the best one can do is
to derive constraints on the fields; for example the
aforementioned attractor equations. And even then, deriving these
equations is a long and difficult process. One may then desire to
find an approach to constructing $D=5$ hypermultiplet solutions
that is more systematic and hopefully easily generalizable to
other types of fields in other dimensions. One way of doing this,
which we propose in this article, is by exploiting the symplectic
nature of the theory. It has long been known that quaternionic and
special K\"{a}hler geometries contain symplectic isometries and
that the hypermultiplets action (with or without gravity) is in
fact symplectically invariant. Furthermore, direct
examination of known constructions reveals that they are written
in terms of symplectic invariants and that this seems to be a
recurrent theme. So the question becomes, can one construct
solutions based solely on symplectic invariance? If so, what is
the simplest form of the theory's field/supersymmetry equations
that reduces the amount of work needed to verify these
ans\"{a}tze? In this paper, this is exactly what we attempt to
explore.

The paper is structured in the following way: Section \ref{modulireview} reviews the definition of the space of complex structure moduli of Calabi-Yau manifolds. In section \ref{SKGandSp} we discuss special K\"{a}hler geometry with
particular emphasis on its symplectic structure. In so doing, we
set the notation needed for dealing with symplectic invariants,
collect all the necessary equations from the literature, as well
as derive new quantities. Section \ref{dimensionalreduction}
reviews the dimensional reduction of $D=11$ SUGRA over a
Calabi-Yau 3-fold with nontrivial complex structure moduli.
Finally, in section \ref{mathematicasymplectica} we put everything
together and reformulate the theory into a symplectically
covariant form and write down the field and SUSY equations in the
simplest way possible. It is our hope that the equations of this
section can be used in future research to straightforwardly write
down and study solution ans\"{a}tze. We conclude by
showing how this approach is applied to two known $D=5$ results.

\section{The space of complex structure moduli of Calabi-Yau manifolds\label{modulireview}}

A Calabi-Yau manifold $\M$ is defined as a K\"{a}hler manifold endowed with Ricci flat metrics. The fields of String/SUGRA theories dimensionally reduced over CY 3-folds generally correspond to the parameters
that describe possible deformations of $\M$. This
parameters' space factorizes, at least locally, into a product
manifold ${\mathcal{M}}_C \otimes {\mathcal{M}}_K$, with
${\mathcal{M}}_C$ being the manifold of complex structure moduli and
${\mathcal{M}}_K$ being a complexification of the parameters of
the K\"{a}hler class. These so-called moduli spaces turn out to
belong to the category of special K\"{a}hler manifolds (defined in the next section).

Calabi-Yau 3-folds admit a single (3,0)
cohomology form; \emph{i.e.} they have Hodge number $h_{3,0}= 1$,
which we will call $\Omega$ (the holomorphic volume form) and an
arbitrary number of (1,1) and (2,1) forms determined by the
corresponding $h$'s (whose values depend on the
particular choice of CY manifold). The Hodge number $h_{2,1}$
determines the dimensions of ${\mathcal{M}}_C$, while $h_{1,1}$
determines the dimensions of ${\mathcal{M}}_K$. The pair
($\mathcal{M},K$), where $K$ is the K\"{a}hler form of $\mathcal{M}$, can be deformed by either deforming
the complex structure of $\mathcal{M}$ or by deforming the
K\"{a}hler form $K$ (or both). In particular, ${\mathcal{M}}_C$ corresponds to special Lagrangian cycles of the
CY space $\M$ that are completely specified by
knowledge of the unique $(3,0)$ form $\Omega$ and the arbitrary
number of $(2,1)$ forms.

The following basic properties of $\Omega$ can be found:
\bea
    \int\limits_\M {\Omega  \wedge \bar \Omega }  &=&  - ie^{ - \K}  \quad\quad\quad\quad
    \int\limits_\M {\Omega  \wedge \nabla _i \Omega }  = \int\limits_\M {\bar \Omega  \wedge \nabla _{\bar i} \bar \Omega }  = 0 \nonumber\\
    \int\limits_\M {\nabla _i \Omega  \wedge \nabla _{\bar j} \bar \Omega }  &=& iG_{i\bar j} e^{ - \K}\quad\quad\quad \left( {i = 1, \ldots ,h_{2,1} } \right),\label{Omegarelations}
\eea
where $\K$ is the K\"{a}hler potential of ${\mathcal{M}}_C$, $G_{i\bar j}$ is a complex metric on ${\mathcal{M}}_C$ and $\nabla$ is defined by
\be
    \nabla _i  = \partial _i  + \frac{1}{2}\left( {\partial _i \K } \right),\,\,\,\,\,\,\,\,\,\,\,\,\,\,\nabla _{\bar i}  = \partial _{\bar i}  - \frac{1}{2}\left( {\partial _{\bar i} \K } \right),
\ee
based on the $U(1)$ K\"{a}hler connection
\be
    \mathcal{P} =  - \frac{i}{2}\left[ {\left( {\partial _i \K} \right)dz^i  - \left( {\partial _{\bar i} \K} \right)dz^{\bar i} }
    \right].\label{U1connection2}
\ee

The space $\mathcal{M}_C$ can be
described in terms of the periods of $\Omega$. Let $\left( {A^I ,B_J } \right)$, where $I,J,K = 0,
\ldots ,h_{2,1} $, be a canonical $H^3$ homology basis such that
\bea
    A^I  \cap B_J  &=& \delta _J^I ,\quad\quad\quad\quad\quad\quad B_I  \cap A^J  =  - \delta _I^J  \nonumber\\
    A^I  \cap A^J  &=&  B_I  \cap B_J  = 0,
\eea
and let $\left( {\alpha _I ,\beta ^J } \right)$ be the dual
cohomology basis forms such that
\begin{eqnarray}
    \int\limits_\M {\alpha _I  \wedge } \beta ^J  &=&  \int\limits_{A^J } {\alpha _I } = \delta
    _I^J,\quad\quad\quad\quad  \int\limits_\M {\beta ^I  \wedge \alpha _J } = \int\limits_{B_J } {\beta ^I } =  -
    \delta _J^I , \nonumber \\
    \int\limits_\M {\alpha _I  \wedge } \alpha _J  &=& \int\limits_\M {\beta ^I  \wedge \beta ^J }  =
    0.
    \label{cohbasis}
\end{eqnarray}

The periods of $\Omega$ are then defined by
\begin{equation}\label{periods}
    Z^I  = \int\limits_{A^I } {\Omega },\quad\quad F_I  = \int\limits_{B_I
    } \Omega,
\end{equation}
such that
\begin{equation}\label{defomega}
    \Omega  = Z^I \alpha _I  - F_I \beta ^I,
\end{equation}
and the K\"{a}hler potential of $\mathcal{M}_C$ becomes
\begin{equation}\label{pot}
    \K =  - \ln \left[ {i\left( {\bar Z^I F_I  - Z^I \bar F_I }
    \right)} \right].
\end{equation}

The so-called periods matrix is defined by
\be
    \N_{IJ} = \bar F_{IJ} +2 i \frac{{N_{IK} Z^K N_{JL} Z^L
    }}{{Z^PN_{PQ} Z^Q }}= \theta_{IJ}-i \gamma_{IJ}\label{gammathetadefined}
\ee
where $F_{IJ}  = \partial _I F_J $ (the derivative is with respect
to $Z^I$), $N_{IJ}=Im(F_{IJ})$ and
$\gamma^{IJ}\gamma_{JK}=\delta^I_K$.

Finally, we note that one can choose a set of
independent ``special coordinates'' $z$ as follows:
\begin{equation}
    z^I  = \frac{{Z^I }}{{Z^0 }},
\end{equation}
which are identified with the moduli of the complex structure $z^i$.

\section{Special geometry and symplectic covariance}\label{SKGandSp}

The space $\mathcal{M}_C$ is described by special K\"{a}hler geometry, which we define in this section. The language we will use relies heavily on the symplectic
structure of special manifolds. Some of the notation and equations
used here are original to this work. Our objective is to develop a
working formulation of symplectic vector spaces that should
facilitate the analysis of solutions in the hypermultiplets sector
of $D=5$ $\N=2$ SUGRA, as well as any other theory with symplectic
structure.

The symplectic group
$Sp\left( {2m,\mathbb{F}} \right) \subset GL\left( {2m,\mathbb{F}}
\right)$ is the isometry group of a nondegenerate alternating
bilinear form on a vector space of rank $2m$ over $\mathbb{F}$,
where this last is usually either $\mathbb{R}$ or $\mathbb{C}$,
although other generalizations are possible. For our purposes, we
take $\mathbb{F}=\mathbb{R}$ and $m=\htwo+1$. In other words,
$Sp\left( {2\htwo+2,\mathbb{R}} \right)$ is the group of the real
bilinear matrices
\be
    {\bf \Lambda } = \left[ {\begin{array}{*{20}c}
    {{}^{ {11} }\Lambda _J^I } &  {{}^{ {12} }\Lambda ^{IJ} } \\
    {{}^{ {21} }\Lambda _{IJ} } & {{}^{ {22} }\Lambda _I^J }  \\
    \end{array}} \right] \in Sp\left( {2\htwo+2,\mathbb{R}} \right)
\ee
that leave the totally antisymmetric symplectic matrix:
\be\label{Sp metric}
    {\bf S} = \left[ {\begin{array}{*{20}c}
   0 & \mathbbm{1}  \\
   { - \mathbbm{1}} & {0}  \\
\end{array}} \right] = \left[ {\begin{array}{*{20}c}
   0 & {\delta _I^J }  \\
   { - \delta _J^I } & 0  \\
\end{array}} \right]
\ee
invariant; \emph{i.e.}
\be\label{Sp condition 1}
    {\bf \Lambda }^T {\bf S\Lambda } = {\bf S}\quad\quad\quad\quad\quad\quad{\bf \Lambda }^T {{\bf S}^T{\bf\Lambda} } = {\bf S}^T,
\ee
implying $\left| {\bf \Lambda } \right| =
\mathbbm{1}$. The inverse of ${\bf \Lambda }$ is found to be:
\be
 {\bf \Lambda }^{ - 1}  = {\bf S}^{ - 1} {\bf \Lambda}^T
 {\bf S}=\left[ {\begin{array}{*{20}c}
    {{}^{ {22} }\Lambda _J^I } &  -{{}^{ {12} }\Lambda ^{IJ} } \\
    -{{}^{ {21} }\Lambda _{IJ} } &  {{}^{ {11} }\Lambda _I^J } \\
    \end{array}} \right],\label{Sp condition 2}
\ee
such that, using (\ref{Sp condition 1}), ${\bf \Lambda }^{ - 1}
{\bf \Lambda } = {\bf S}^{ - 1} {\bf \Lambda }^T {\bf S\Lambda } =
{\bf S}^{ - 1} {\bf S} = \mathbbm{1}$ as needed. Also note that
${\bf S}^{ - 1}  = {\bf S}^T  = - {\bf S}$. We adopt the language
that there exists a vector space \textbf{\textit{Sp}} such that
the symplectic matrix $\bf S$ acts as a metric on that space.
Symplectic vectors in \textbf{\textit{Sp}} can be written in a
``ket'' notation as follows
\be
    \left| A \right\rangle  = \left( {\begin{array}{*{20}c}
   {a^I }  \\
   {\tilde a_I }  \\
    \end{array}} \right),\quad \left| B \right\rangle  = \left( {\begin{array}{*{20}c}
   {b^I }  \\
   {\tilde b_I }  \\
    \end{array}} \right).
\ee

On the other hand, ``bra'' vectors defining a space dual to
\textbf{\textit{Sp}} can be found by contraction with the metric
in the usual way, yielding:
\be
    \left\langle A \right| = \left( {{\bf SA}} \right)^T  = {\bf
    A}^T {\bf S}^T  = \begin{array}{*{20}c}
   {\left( {\begin{array}{*{20}c}
   {a^J } & {\tilde a_J }  \\
    \end{array}} \right)}  \\
   {}  \\
    \end{array}\left[ {\begin{array}{*{20}c}
   0 & { - \delta _J^I }  \\
   {\delta _I^J } & 0  \\
    \end{array}} \right] = \begin{array}{*{20}c}
   {\left( {\begin{array}{*{20}c}
   {\tilde a_I } & { - a^I }  \\
    \end{array}} \right)}  \\
   {}  \\
    \end{array},
\ee
such that the inner product on \textbf{\textit{Sp}} is the
``bra(c)ket'':
\be
    \left\langle {A}
 \mathrel{\left | {\vphantom {A B}}
 \right. \kern-\nulldelimiterspace}
 {B} \right\rangle  = {\bf A}^T {\bf S}^T {\bf B} = \begin{array}{*{20}c}
   {\left( {\begin{array}{*{20}c}
   {\tilde a_I } & { - a^I }  \\
\end{array}} \right)}  \\
   {}  \\
\end{array}\left( {\begin{array}{*{20}c}
   {b^I }  \\
   {\tilde b_I }  \\
\end{array}} \right) = \tilde a_I b^I  - a^I \tilde b_I  =  - \left\langle {B}
 \mathrel{\left | {\vphantom {B A}}
 \right. \kern-\nulldelimiterspace}
 {A} \right\rangle.\label{Sp inner product}
\ee

In this language, the matrix ${\bf \Lambda }$ can simply be
thought of as a rotation operator in \textbf{\textit{Sp}}. So a
rotated vector is
\be\label{SpRotation}
    \left| {A'} \right\rangle  =  \pm \left|\Lambda A \right\rangle  = \pm {\bf \Lambda A}.
\ee

This is easily shown to preserve the inner product (\ref{Sp inner
product}):
\be
    \left\langle {{A'}}
 \mathrel{\left | {\vphantom {{A'} {B'}}}
 \right. \kern-\nulldelimiterspace}
 {{B'}} \right\rangle  = \left(  \pm  \right)^2 {\bf A}^T {\bf \Lambda }^T {\bf S}^T {\bf \Lambda B} = {\bf A}^T {\bf S}^T {\bf B} = \left\langle {A}
 \mathrel{\left | {\vphantom {A B}}
 \right. \kern-\nulldelimiterspace}
    {B} \right\rangle,
\ee
where (\ref{Sp condition 1}) was used. In fact, one can
\emph{define} (\ref{Sp condition 1}) based on the requirement that
the inner product is preserved. To facilitate future calculations,
we define the symplectic invariant
\bea
    \left\langle A \right|\Lambda \left| B \right\rangle  &\equiv& \left\langle {A}
 \mathrel{\left | {\vphantom {A {\Lambda B}}}
 \right. \kern-\nulldelimiterspace}
 {{\Lambda B}} \right\rangle  =  {\bf A}^T {\bf S}^T {\bf \Lambda B}\nonumber\\
 &=&  \left\langle {{A\Lambda ^{ - 1} }}
 \mathrel{\left | {\vphantom {{A\Lambda ^{ - 1} } B}}
 \right. \kern-\nulldelimiterspace}
 {B} \right\rangle=- \left\langle {{ B\Lambda}}
 \mathrel{\left | {\vphantom {{ B\Lambda} A}}
 \right. \kern-\nulldelimiterspace}
 {A} \right\rangle.\label{SpInvariant}
\eea

The matrix $\bf\Lambda$ we will be using in the remainder of the
paper has the property
\be
   {}^{22}\Lambda _J^I  =  - {}^{11}\Lambda _J^I \quad  \to \quad {\bf \Lambda }^{ - 1}  =  - {\bf \Lambda
   },\label{LambdaProperty}
\ee
which, via (\ref{SpInvariant}), leads to
\be
    \left\langle A \right|\Lambda \left| B \right\rangle  = \left\langle {A}
 \mathrel{\left | {\vphantom {A {\Lambda B}}}
 \right. \kern-\nulldelimiterspace}
 {{\Lambda B}} \right\rangle  =  - \left\langle {{A\Lambda }}
 \mathrel{\left | {\vphantom {{A\Lambda } B}}
 \right. \kern-\nulldelimiterspace}
     {B} \right\rangle.
\ee

The choice (\ref{LambdaProperty}) is not the only natural one. A
consequence of it is that $\bf \Lambda$ is not symmetric, but
${\bf S \Lambda}$ is. On the other hand an equivalent choice would
be a symmetric $\bf\Lambda$, in which case it would be ${\bf S
\Lambda}$ that satisfies (\ref{LambdaProperty}). Within the
context of special geometry, we have opted for a nonsymmetric
$\bf\Lambda$ since it makes some later equations simpler.

Now consider the algebraic product of the two symplectic scalars
\be\label{inter1}
    \left\langle {A}
 \mathrel{\left | {\vphantom {A B}}
 \right. \kern-\nulldelimiterspace}
 {B} \right\rangle \left\langle {C}
 \mathrel{\left | {\vphantom {C D}}
 \right. \kern-\nulldelimiterspace}
    {D} \right\rangle  = \left( {{\bf A}^T {\bf S}^T {\bf B}} \right)\left( {{\bf C}^T {\bf S}^T {\bf D}} \right).
\ee

The ordinary outer product of matrices is defined by
\be
    {\bf B} \otimes {\bf C}^T  = \left( {\begin{array}{*{20}c}
   {b^I }  \\
   {\tilde b_I }  \\
\end{array}} \right)\begin{array}{*{20}c}
   { \otimes \left( {\begin{array}{*{20}c}
   {c^J } & {\tilde c_J }  \\
\end{array}} \right)}  \\
   {}  \\
\end{array} = \left[ {\begin{array}{*{20}c}
   {b^I c^J } & {b^I \tilde c_J }  \\
   {\tilde b_I c^J } & {\tilde b_I \tilde c_J }  \\
    \end{array}} \right],
\ee
which allows us to rewrite (\ref{inter1}):
\be
    \left\langle {A}
 \mathrel{\left | {\vphantom {A B}}
 \right. \kern-\nulldelimiterspace}
 {B} \right\rangle \left\langle {C}
 \mathrel{\left | {\vphantom {C D}}
 \right. \kern-\nulldelimiterspace}
    {D} \right\rangle  = {\bf A}^T {\bf S}^T \left( {{\bf B} \otimes {\bf C}^T {\bf S}^T } \right){\bf D} = \left\langle A \right|{\bf B} \otimes {\bf C}^T {\bf S}^T \left| D \right\rangle.\label{Inter2}
\ee

Comparing the terms of (\ref{Inter2}), we conclude that one way a
symplectic outer product can be defined is:
\be
    \left| B \right\rangle \left\langle C \right| = {\bf B} \otimes {\bf C}^T {\bf S}^T  = \left[ {\begin{array}{*{20}c}
   {b^I\tilde c_J} & { - b^I c^J}  \\
   {\tilde b_I\tilde c_J} & { - \tilde b_I c^J}  \\
    \end{array}} \right].\label{Spouterproduct}
\ee

Note that the order of vectors in (\ref{Spouterproduct}) is
important, since generally
\be
    \left| B \right\rangle \left\langle C \right| = \left[ {{\bf
    S}\left| C \right\rangle \left\langle B \right|{\bf S}} \right]^T.
\ee

However, if the outer product $\left| B \right\rangle \left\langle
C \right|$ satisfies the property (\ref{LambdaProperty}),
\emph{i.e.}
\be
    \left[ {\left| B \right\rangle \left\langle C \right|} \right]^{ - 1}  =  - \left| B \right\rangle \left\langle
    C \right|,
\ee
then it is invariant under the interchange $B \leftrightarrow C$:
\be
    \left| B \right\rangle \left\langle C \right| = \left| C \right\rangle \left\langle B
    \right|.
\ee

The definition of a special K\"{a}hler manifold goes like this: Let $\mathcal{L}$
denote a complex $U(1)$ line bundle whose first Chern class equals the
K\"{a}hler form $\K$ of a Hodge-K\"{a}hler manifold $\M$. Now
consider an additional holomorphic flat vector bundle of rank
$(2\htwo+2)$ with structural group $Sp(2\htwo+2, \mathbb{R})$ on
$\M$: $\mathcal{SV}\rightarrow \M$. Construct a tensor bundle
$\mathcal{SV}\otimes \mathcal{L}$. This then is a
special K\"{a}hler manifold if for some holomorphic section
$\left| \Psi  \right\rangle $ of such a bundle the K\"{a}hler
2-form is given by:
\be
    K = -\frac{i}{{2\pi }}\partial \bar \partial \ln \left( {i\left\langle {\Psi }
 \mathrel{\left | {\vphantom {\Psi  { \bar\Psi }}}
 \right. \kern-\nulldelimiterspace}
 {{ \bar\Psi }} \right\rangle } \right),
\ee
or in terms of the K\"{a}hler potential:
\be
    \K =  - \ln \left( {i\left\langle {\Psi }
 \mathrel{\left | {\vphantom {\Psi  { \bar\Psi }}}
 \right. \kern-\nulldelimiterspace}
 {{ \bar\Psi }} \right\rangle } \right)\quad  \to \quad \left\langle {\bar\Psi }
 \mathrel{\left | {\vphantom {\bar\Psi  { \Psi }}}
 \right. \kern-\nulldelimiterspace}
 {{ \Psi }} \right\rangle  = ie^{ - \K}.\label{pot1} \ee

Now, this exactly describes the space
of complex structure moduli $\M_C$ if one chooses:
\be
    \left| \Psi  \right\rangle  = \left( {\begin{array}{*{20}c}
   {Z^I }  \\
   {F_I }  \\
    \end{array}} \right), \label{periodvector}
\ee
which, via (\ref{pot1}), leads directly to equation (\ref{pot})
defining the K\"{a}hler potential of $\M_C$. We then identify
$\M_C$ as a special K\"{a}hler manifold with metric $G_{i \bar
j}$.

It can be easily demonstrated that the matrix:
\be\label{symplecticmatrix1}
    {\bf \Lambda } = \left[ {\begin{array}{*{20}c}
    {\gamma ^{IK} \theta _{KJ} } & -{\gamma ^{IJ} } \\
    {\left( {\gamma _{IJ}  + \gamma ^{KL} \theta _{IK} \theta _{JL} } \right)} &  - {\gamma ^{JK} \theta _{KI} } \\
    \end{array}} \right]
\ee
satisfies the symplectic condition (\ref{Sp condition 1}), where
$\gamma$ and $\theta$ are defined by (\ref{gammathetadefined}).
Its inverse is then
\be
    {\bf \Lambda }^{ - 1}  =-\bf \Lambda= \left[ {\begin{array}{*{20}c}
    - {\gamma ^{JK} \theta _{KI} } & {\gamma ^{IJ} }  \\
    -{\left( {\gamma _{IJ}  + \gamma ^{KL} \theta _{IK} \theta _{JL} } \right)} &  {\gamma ^{IK} \theta _{KJ} } \\
    \end{array}} \right].
\ee

The symplectic structure manifest here is a consequence of the
topology of the Calabi-Yau manifold $\M$, the origins of which can
be traced to the completeness relations (\ref{cohbasis}), clearly:
\be
   \int\limits_\M {\left[ {\begin{array}{*{20}c}
   {\alpha _I  \wedge \alpha _J } & {\alpha _I  \wedge \beta ^J }  \\
   {\beta ^I  \wedge \alpha _J } & {\beta ^I  \wedge \beta ^J }  \\
    \end{array}} \right]}= \left[ {\begin{array}{*{20}c}
   0 & {\delta _I^J }  \\
   { - \delta _J^I } & 0  \\
    \end{array}} \right] = {\bf S}.
\ee

In fact, if one defines the symplectic vector:
\be
    \left| \Theta  \right\rangle  = \left( {\begin{array}{*{20}c}
   {\beta ^I }  \\
   {\alpha _I }  \\
    \end{array}} \right),
\ee
then it is easy to check that
\be
    \int\limits_\M {{\bf \Theta }\mathop  \otimes \limits_ \wedge  {\bf \Theta }^T }  = {\bf S}^T \quad  \to \quad \int\limits_\M {\left| \Theta  \right\rangle \mathop  \wedge  \left\langle \Theta  \right|}  =  - \mathbbm{1}.
\ee

Next, we construct a basis in \textbf{\textit{Sp}}. Properly
normalized, the periods vector (\ref{periodvector}) provides such
a basis:
 \be
    \left| V \right\rangle  = e^{\frac{\K}{2}} \left| \Psi  \right\rangle  = \left( {\begin{array}{*{20}c}
   {L^I }  \\
   {M_I }  \\
    \end{array}} \right),
\ee
such that, using (\ref{pot1}):
\be
    \left\langle {{\bar V}}
 \mathrel{\left | {\vphantom {{\bar V} V}}
 \right. \kern-\nulldelimiterspace}
 {V} \right\rangle  = \left( {L^I \bar M_I  - \bar L^I M_I } \right) =
 i.\label{SpBasisNorm}
\ee

Since ${\left| V \right\rangle }$ is a scalar in the
$\left(i,j,k\right)$ indices, it couples only to the
$U\left(1\right)$ bundle via the K\"{a}hler covariant derivative:
\bea
     \left|\nabla _i V \right\rangle  &=& \left|\left[ {\partial _i  + \frac{1}{2}\left( {\partial _i \K} \right)} \right] V \right\rangle ,\quad \quad  \left|\nabla _{\bar i} V \right\rangle  =\left| \left[ {\partial _{\bar i}  - \frac{1}{2}\left( {\partial _{\bar i} \K} \right)} \right] V \right\rangle  \nonumber\\
     \left|\nabla _i {\bar V} \right\rangle  &=& \left|\left[ {\partial _i  - \frac{1}{2}\left( {\partial _i \K} \right)} \right] {\bar V} \right\rangle ,\quad \quad  \left|\nabla _{\bar i} {\bar V} \right\rangle  = \left|\left[ {\partial _{\bar i}  + \frac{1}{2}\left( {\partial _{\bar i} \K} \right)} \right] {\bar V}
    \right\rangle.
\eea

Using this, one can construct the
orthogonal \textbf{\textit{Sp}} vectors:
\bea
    \left| {U_i } \right\rangle  &=& \left| \nabla _i V
    \right\rangle  =  \left(
    {\begin{array}{*{20}c}
   {\nabla _i L^I }  \\
   {\nabla _i M_I }  \\
    \end{array}} \right) = \left( {\begin{array}{*{20}c}
   {f_i^I }  \\
   {h_{i|I} }  \\
    \end{array}} \right) \\
     \left| {U_{\bar i} } \right\rangle  &=& \left|\nabla _{\bar i}  {\bar V} \right\rangle  =\left( {\begin{array}{*{20}c}
   {\nabla _{\bar i} \bar L^I }  \\
   {\nabla _{\bar i} \bar M_I }  \\
    \end{array}} \right) = \left( {\begin{array}{*{20}c}
   {f_{\bar i}^I }  \\
   {h_{\bar i|I} }  \\
    \end{array}} \right),
\eea
with
\bea
     \left|\nabla _i U_j  \right\rangle  &=& \left|\left[ {\partial _i  + \frac{1}{2}\left( {\partial _i \K} \right)} \right] U_j  \right\rangle ,\quad \quad  \left|\nabla _{\bar i} U_j  \right\rangle  =\left| \left[ {\partial _{\bar i}  - \frac{1}{2}\left( {\partial _{\bar i} \K} \right)} \right] U_j  \right\rangle  \nonumber\\
     \left|\nabla _i U_{\bar j} \right\rangle  &=& \left|\left[ {\partial _i  - \frac{1}{2}\left( {\partial _i \K} \right)} \right] U_{\bar j} \right\rangle ,\quad \quad  \left|\nabla _{\bar i} U_{\bar j} \right\rangle  = \left|\left[ {\partial _{\bar i}  + \frac{1}{2}\left( {\partial _{\bar i} \K} \right)} \right] U_{\bar j}
    \right\rangle.
\eea

Note that $\left| {U_i } \right\rangle$ also couples to the metric
$G_{i\bar j}$ via the Levi-Civita connection. So its full
covariant derivative is defined by:
\bea
    \left| {\D_i U_j } \right\rangle  &=& \left| {\nabla _i U_j } \right\rangle  - \Gamma _{ij}^k \left| {U_k } \right\rangle \quad \quad \left| {\D_{\bar i} U_j } \right\rangle  = \left| {\nabla _{\bar i} U_j } \right\rangle  \nonumber\\
    \left| {\D_i U_{\bar j} } \right\rangle  &=& \left| {\nabla _i U_{\bar j} } \right\rangle \quad \quad \quad\quad\quad\quad\;\left| {\D_{\bar i} U_{\bar j} } \right\rangle  = \left| {\nabla _{\bar i} U_{\bar j} } \right\rangle  - \Gamma _{\bar i\bar j}^{\bar k} \left| {U_{\bar k} }
    \right\rangle.
\eea

It can be demonstrated that these quantities satisfy the
properties
\bea
    \left|\nabla _i  {\bar V} \right\rangle  &=& \left|\nabla _{\bar i}  V \right\rangle =0\label{Normality2}\\
    \left\langle {{U_i }}
    \mathrel{\left | {\vphantom {{U_i } {U_j }}}
    \right. \kern-\nulldelimiterspace}
    {{U_j }} \right\rangle  &=& \left\langle {{U_{\bar i} }}
    \mathrel{\left | {\vphantom {{U_{\bar i} } {U_{\bar j} }}}
    \right. \kern-\nulldelimiterspace}
    {{U_{\bar j} }} \right\rangle    =0\\
    \left\langle {\bar V}
    \mathrel{\left | {\vphantom {\bar V {U_i }}}
    \right. \kern-\nulldelimiterspace}
    {{U_i }} \right\rangle  &=& \left\langle {V}
    \mathrel{\left | {\vphantom {V {U_{\bar i} }}}
    \right. \kern-\nulldelimiterspace}
    {{U_{\bar i} }} \right\rangle  = \left\langle { V}
    \mathrel{\left | {\vphantom { V {U_i }}}
    \right. \kern-\nulldelimiterspace}
    {{U_i }} \right\rangle=\left\langle {\bar V}
    \mathrel{\left | {\vphantom {\bar V {U_{\bar i} }}}
    \right. \kern-\nulldelimiterspace}
    {{U_{\bar i} }} \right\rangle= 0,\label{Normality}\\
    \left|\nabla _{\bar j}  {U_i } \right\rangle  &=& G_{i\bar j} \left| V \right\rangle \nonumber\\ \left|\nabla _i  {U_{\bar j} } \right\rangle  &=& G_{i\bar j} \left| {\bar V}
    \right\rangle,\\
    G_{i\bar j}&=& \left( {\partial _i \partial _{\bar j} \K} \right)=- i    \left\langle {{U_i }}
    \mathrel{\left | {\vphantom {{U_i } {U_{\bar j} }}}
    \right. \kern-\nulldelimiterspace}
    {{U_{\bar j} }} \right\rangle.\label{KmetricasSpproduct}
\eea

Special K\"{a}hler manifolds admit a completely
symmetric and covariantly holomorphic tensor $C_{ijk}$ and its
antiholomorphic conjugate $C_{\bar i\bar j\bar k}$ such that the
following restriction on the curvature is true:
\be
    R_{\bar i j\bar k l}  = G_{j\bar k} G_{l\bar i}  + G_{l\bar k} G_{j\bar i}  - C_{rlj} C_{\bar s\bar i\bar k} G^{r\bar
    s},\label{specialcurvature}
\ee
generally referred to in the literature as the special
K\"{a}hler geometry constraint. It can be shown that
\be
    \left| {\D _i U_j } \right\rangle  = G^{k\bar l} C_{ijk} \left| {U_{\bar l} }
    \right\rangle,
\ee
which leads to:
\be
    C_{ijk}  = -i\left\langle {{\D _i U_j }}
    \mathrel{\left | {\vphantom {{\D _i U_j } {U_k }}}
    \right. \kern-\nulldelimiterspace}
    {{U_k }} \right\rangle.
\ee

The following identities may now be derived:
\bea
    \mathcal{N}_{IJ} L^J  &=& M_I ,\quad \quad \quad \quad \mathcal{\bar N}_{IJ} f_i^J  = h_{i|I}  \nonumber\\
    \mathcal{\bar N}_{IJ} {\bar L}^J  &=& {\bar M}_I ,\quad \quad \quad \quad \mathcal{ N}_{IJ}  f_{\bar i}^J  =  h_{{\bar i}|I}  \label{PeriodMatrix}\\
    \gamma _{IJ} L^I \bar L^J  &=& \frac{1}{2},\quad \quad \quad \quad G_{i\bar j}  = 2\gamma _{IJ} f_i^I f_{\bar j}^J,  \label{gammametric}
\eea
as well as the very useful (and quite essential for our purposes)
\bea
    \gamma ^{IJ}  &=& 2\left( L^I \bar L^J + {G^{i\bar j} f_i^I f_{\bar j}^J   } \right)  \nonumber\\
    \left( {\gamma _{IJ}  + \gamma ^{KL} \theta _{IK} \theta _{JL} } \right) &=& 2\left( { M_I \bar M_J + G^{i\bar j} h_{i|I} h_{\bar j|J} } \right) \nonumber\\
    \gamma ^{IK} \theta _{KJ}&=&2\left(\bar L^I M_J + G^{i\bar j}f_i^{\,\,I}h_{\bar
    j|J}\right)+i \delta^I_J\nonumber\\
    &=&2\left(L^I \bar M_J + G^{i\bar j}h_{i|J} f_{\bar j}^{\,\,\,I}\right)-i\delta^I_J\nonumber\\
    &=& \left({L^I \bar M_J  + \bar L^I M_J } \right) + G^{i\bar j} \left( {f_i^{\,\,I}h_{\bar j|J}  + h_{i|J} f_{\bar j}^{\,\,\,I} } \right).\label{GammaThetaLMconnection}
\eea

Equations (\ref{GammaThetaLMconnection}) lead to a
second form for the symplectic matrix (\ref{symplecticmatrix1}):
\be\label{symplecticmatrix2}
    {\bf \Lambda } = \left[ {\begin{array}{*{20}c}
        {\left({L^I \bar M_J  + \bar L^I M_J } \right)} & {} & {-{2\left( {L^I \bar L^J+G^{i\bar j} f_i^I f_{\bar j}^J   } \right)}}  \\
        {+{G^{i\bar j} \left( {f_i^{\,\,I}h_{\bar j|J}  + h_{i|J} f_{\bar j}^{\,\,\,I} } \right)}} & {}  \\
        {} & {} & {-\left({L^J \bar M_I  + \bar L^J M_I } \right)}  \\
        {2\left( {M_I \bar M_J+G^{i\bar j} h_{i|I} h_{\bar j|J}   } \right)} & {} & {-G^{i\bar j} \left( {f_i^{\,\,J}h_{\bar j|I}  + h_{i|I} f_{\bar j}^{\,\,\,J} } \right)}  \\
    \end{array}} \right]
\ee
with inverse
\be
    {\bf \Lambda }^{ - 1}  = -{\bf \Lambda } =\left[ {\begin{array}{*{20}c}
        {-\left({L^J \bar M_I  + \bar L^J M_I } \right)} & {} & {{2\left( {L^I \bar L^J+G^{i\bar j} f_i^I f_{\bar j}^J   } \right)}}  \\
        {-G^{i\bar j} \left( {f_i^{\,\,J}h_{\bar j|I}  + h_{i|I} f_{\bar j}^{\,\,\,J} } \right)} & {} & {}  \\
        {} & {} & {\left({L^I \bar M_J  + \bar L^I M_J } \right)}  \\
        {-2\left( {M_I \bar M_J +G^{i\bar j} h_{i|I} h_{\bar j|J}  } \right)} & {} & {+{G^{i\bar j} \left( {f_i^{\,\,I}h_{\bar j|J}  + h_{i|J} f_{\bar j}^{\,\,\,I} } \right)}}  \\
    \end{array}} \right].
\ee

By inspection, one can write down the following important result:
\bea
    {\bf \Lambda } &=& \left| V \right\rangle \left\langle {\bar V} \right| + \left| {\bar V} \right\rangle \left\langle V \right| + G^{i\bar j} \left| {U_i } \right\rangle \left\langle {U_{\bar j} } \right| + G^{i\bar j} \left| {U_{\bar j} } \right\rangle \left\langle {U_i } \right| \nonumber\\
    {\bf \Lambda }^{ - 1}  &=&  - \left| V \right\rangle \left\langle {\bar V} \right| - \left| {\bar V} \right\rangle \left\langle V \right| - G^{i\bar j} \left| {U_i } \right\rangle \left\langle {U_{\bar j} } \right| - G^{i\bar j} \left| {U_{\bar j} } \right\rangle \left\langle {U_i } \right|.\label{Lambdaasouter1}
\eea

In other words, the rotation matrix in \textbf{\textit{Sp}} is
expressible as the outer product of the basis vectors; a result
which, in retrospect, seems obvious. Note that since $\bf \Lambda$
satisfies the property (\ref{LambdaProperty}), it is invariant
under the interchange $V \leftrightarrow \bar V$ and/or $U_i
\leftrightarrow U_{\bar j}$. This makes manifest the fact that
$\bf \Lambda$ is a real matrix; ${\bf \Lambda } = {\bf \bar
\Lambda }$. Now, applying ${\bf \Lambda }^{ - 1} {\bf \Lambda } =
\mathbbm{1}$, we end up with the condition
\be
    \left| {\bar V} \right\rangle \left\langle V \right| + G^{i\bar j} \left| {U_i } \right\rangle \left\langle {U_{\bar j} } \right|=\left| V \right\rangle \left\langle {\bar V} \right|+G^{i\bar j} \left| {U_{\bar j} } \right\rangle \left\langle {U_i } \right|-i,
\ee
which can be checked explicitly using
(\ref{GammaThetaLMconnection}). This can be used to write $\bf
\Lambda$ in an even simpler form:
\bea
    {\bf \Lambda } &=& 2\left| V \right\rangle \left\langle {\bar V} \right| + 2G^{i\bar j} \left| {U_{\bar j} } \right\rangle \left\langle {U_i } \right|
    -i\nonumber\\
    {\bf \Lambda }^{-1} &=& -2\left| V \right\rangle \left\langle {\bar V} \right| - 2G^{i\bar j} \left| {U_{\bar j} } \right\rangle \left\langle {U_i } \right|
    +i.\label{Lambdaasouter2}
\eea

For future convenience we also compute
\bea
    \D_i {\bf \Lambda } =\nabla_i {\bf \Lambda }= \partial_i {\bf \Lambda } = 2\left| {U_i } \right\rangle \left\langle {\bar V} \right|+2\left| {\bar V} \right\rangle \left\langle {U_i } \right| + 2G^{j\bar r} G^{k\bar p} C_{ijk} \left| {U_{\bar r} } \right\rangle \left\langle {U_{\bar p} } \right|.
    \label{CovariantDerofLambda}
\eea

It is clearly easier, and
possibly more intuitive, to work with an expression such as
(\ref{Lambdaasouter2}) over something like
(\ref{symplecticmatrix2}), or even (\ref{symplecticmatrix1}). It
is indeed this very fact that has motivated this work in its
entirety. Finally, we note that our discussion here is based on a
definition of special manifolds that is not the only one in
existence. See, for instance, \cite{Craps:1997gp} for details.
Explicit examples of special manifolds in various dimensions are
given in, for example, \cite{hep-th/9512043}. More detail on this obviously vast topic may be found in \cite{Joyce, hep-th/9605032, Candelas:1985hv, hep-th/9506150, Ferrara:1991na, Candelas:1989bb, hep-th/9508001,  hep-th/0203247, Emam:2004nc}.

\section{$D=5$ $\N=2$ supergravity with hypermultiplets\label{dimensionalreduction}}

The dimensional reduction of $D=11$ supergravity over a
Calabi-Yau manifold $\M$ yields ungauged $D=5$ $\N=2$
SUGRA. We look at the case where
only the complex structure of $\M$ is deformed. We will follow,
and slightly extend, the notation of \cite{Gutperle:2000ve}.

The unique supersymmetric gravity theory in eleven dimensions has
the following bosonic action:
\be
    S_{11}  = \int_{11} \left( { {\mathcal{R}\star 1 -
    \frac{1}{2}\F \wedge \star \F -
    \frac{1}{6}\A \wedge \F \wedge
    \F}}\right),
\ee
where $\mathcal{R}$ is the $D=11$ Ricci scalar, $\A$ is the 3-form
gauge potential, $\F=d\A$ and $\star$ is the Hodge star operator. The dimensional reduction is
traditionally done using the metric:
\be
    ds^2   = e^{\frac{2}{3}\sigma } g_{\mu \nu } dx^\mu  dx^\nu
    + e^{ - \frac{\sigma }{3}} ds_{CY}^2 \quad
    \quad  \mu ,\nu  = 0, \ldots ,4,\label{expmet}
\ee
where $g_{\mu \nu }$ is the target five dimensional metric,
$ds_{CY}^2$ is a metric on the six dimensional compact subspace $\M$,
the dilaton $\sigma$ is a function in $x^\mu$ only and the warp
factors are chosen to give the conventional numerical coefficients
in five dimensions.

The flux compactification of the gauge field is done by
expanding $\A$ into two forms, one is the five dimensional gauge
field $A$ while the other contains the components of $\A$ on $\M$
written in terms of the cohomology forms $\left(
\alpha_I,\beta^I\right)$ as follows:
\bea
    \A &=& A + \sqrt 2 \left( {\zeta ^I \alpha _I  + \tilde \zeta _I \beta ^I }
    \right),\nonumber\\
    \F &=& d\A = F + \sqrt 2 \left[ {\left( {\partial _\mu \zeta ^I } \right)\alpha _I  + \left( {\partial _\mu \tilde \zeta _I } \right)\beta ^I } \right]\wedge dx^\mu. \label{FExpanded}
\eea

Because of the eleven dimensional Chern-Simons term, the
coefficients $\zeta^I$ and $\tilde \zeta_I$ appear as
pseudoscalar axion fields in the lower dimensional theory. We
also note that $A$ in five dimensions is dual to a scalar field
which we will call $a$ (known as the universal axion). The set
($a$, $\sigma$, $\zeta^0$, $\tilde \zeta_0$) is known as the
universal hypermultiplet\footnote{So-called because it appears in
all Calabi-Yau compactifications, irrespective of the detailed
structure of the CY manifold. We recall that the dilaton $\sigma$
is proportional to the natural logarithm of the volume of $\M$.}.
The rest of the hypermultiplets are ($z^i$, $z^ {\bar i}$,
$\zeta^i$, $\tilde \zeta_i$), where we
recognize the $z$'s as the CY's complex structure moduli.
Note that the total number of scalar fields in the hypermultiplets
sector is $4(h_{2,1}+1)$ (each hypermultiplet has 4 real scalar
fields) which comprises a quaternionic manifold as noted earlier.
Also included in the hypermultiplets are the fermionic partners of
the hypermultiplet scalars known as the hyperini (singular:
hyperino).

The bosonic action of the ungauged five dimensional $\N=2$ supergravity
theory with vanishing vector multiplets is:
\bea
  S_5  &=& \int\limits_5 \left\{ {{R\star 1 - \frac{1}{2}d\sigma \wedge\star d\sigma  - G_{i\bar j} dz^i \wedge\star dz^{\bar j}  - F \wedge \left( {\zeta^I d\tilde \zeta_I  - \tilde \zeta_I d\zeta^I } \right)} - \frac{1}{2}e^{ - 2\sigma } F \wedge \star F} \right. \nonumber\\
  &-&  \left. {e^\sigma  \left[ {\left( {\gamma_{IJ}  + \gamma ^{ KL} \theta _{IK}\theta _{JL} } \right) {d\zeta^I } \wedge\star {d\zeta^J }  + \gamma ^{IJ}  {d\tilde \zeta_I } \wedge\star {d\tilde \zeta_J }  + 2\gamma ^{ IK} \theta_{JK}  {d\zeta^J } \wedge\star {d\tilde \zeta_I } }
  \right]} \right\}. \label{Fullaction}
\eea

Variation of the action gives the following field
equations for $\sigma$, $\left(z^i,z^{\bar i}\right)$, $A$ and
$\left(\zeta^I,\tilde\zeta_I\right)$:
\bea
    \left( {\Delta \sigma } \right)\star 1 - e^\sigma  X + e^{ - 2\sigma } F \wedge \star F &=& 0\label{dilatoneom}\\
    \left( {\Delta z^i } \right)\star 1 + \Gamma _{jk}^i dz^j  \wedge \star dz^k  - \frac{1}{2}e^\sigma  G^{i\bar j} \left( {\partial _{\bar j} X} \right)\star 1 &=& 0 \nonumber\\
    \left( {\Delta z^{\bar i} } \right)\star 1 + \Gamma _{\bar j\bar k}^{\bar i} dz^{\bar j}  \wedge \star dz^{\bar k}  - \frac{1}{2}e^\sigma  G^{\bar ij} \left( {\partial _j X} \right)\star 1 &=& 0\label{zeom} \\
    d^{\dag} \left[ {e^{ - 2\sigma } F + \star\left( {\zeta ^I d\tilde \zeta _I  - \tilde \zeta _I d\zeta ^I } \right)} \right] &=& 0\label{Feomgeneral}\\
    d^\dag\left[ e^\sigma  \gamma ^{ IK} \theta_{JK}   {d  \zeta^J }  + e^\sigma  \gamma ^{ IJ}
    {d  \tilde \zeta_J }+ \zeta^I \star  F\right]&=&0\nonumber \\
    d^\dag\left[ e^\sigma  \left( {\gamma_{IJ}  + \gamma ^{ KL} \theta _{IK}\theta _{JL} } \right) {d  \zeta^J }  + e^\sigma \gamma ^{ JK} \theta_{IK}  {d  \tilde \zeta_J }   - \tilde \zeta_I \star  F
    \right]&=&0,\label{xieom}
\eea
where $d^\dag$ is the adjoint exterior derivative and $\Delta$ is the Laplace de-Rahm operator. For compactness we have defined
\be
    X= {\left( {\gamma_{IJ}  + \gamma ^{ KL} \theta _{IK}\theta _{JL} } \right) {d\zeta^I } \wedge\star {d\zeta^J }  + \gamma ^{IJ}  {d\tilde \zeta_I } \wedge\star {d\tilde \zeta_J }  + 2\gamma ^{ IK} \theta_{JK}  {d\zeta^J } \wedge\star {d\tilde \zeta_I } }
  ,\label{X}
\ee
as well as used the Bianchi identity $d F=0$ to get the given
form of (\ref{xieom}). From a five dimensional perspective, the
moduli $\left(z^i,z^{\bar i}\right)$ behave as scalar fields. We
recall, however, that the behavior of the other fields is
dependent on the moduli, \emph{i.e.} they are functions in them.
Hence it is possible to treat (\ref{zeom}) as constraints that can
be used to reduce the degrees of freedom of the other field
equations. Certain assumptions, however, are needed to perform
this, so we will not do so here since our objective is to discuss
the field equations in their most general form. This is more
properly done in the context of specific solution ans\"{a}tze.

Equations (\ref{Feomgeneral}) and (\ref{xieom}) are clearly the
statements that the forms:
\bea
    \mathcal{J}_2  &=& e^{ - 2\sigma } F + \star\left( {\zeta ^I d\tilde \zeta _I  - \tilde \zeta _I d\zeta ^I } \right)\nonumber\\
    \mathcal{J}_5^I  &=&  e^\sigma  \gamma ^{ IK} \theta_{JK}   {d  \zeta^J }  + e^\sigma  \gamma ^{ IJ}
    {d  \tilde \zeta_J }+ \zeta^I\star F\nonumber\\
    \mathcal{\tilde J}_{5|I}  &=& e^\sigma  \left( {\gamma_{IJ}  + \gamma ^{ KL} \theta _{IK}\theta _{JL} } \right) {d  \zeta^J }  + e^\sigma \gamma ^{ JK} \theta_{IK}  {d  \tilde \zeta_J }   -\tilde \zeta_I \star  F\label{Currents}
\eea
are conserved. These are, in fact, Noether currents corresponding
to certain isometries of the quaternionic manifold defined by the
hypermultiplets as discussed in various sources
\cite{Ferrara:1989ik, Cecotti:1988qn}. From a five dimensional
perspective, they can be thought of as the result of the
invariance of the action under particular infinitesimal shifts of
$A$ and $\left(\zeta, \tilde\zeta\right)$
\cite{Gutperle:2000ve,Gutperle:2000sb}. The charge densities
corresponding to them can then be found in the usual way by:
\be
 \mathcal{Q}_2  = \int {\mathcal{J}_2 },\quad \quad \quad
 \mathcal{Q}_5^I  = \int { \mathcal{J}_5^I} ,\quad \quad \quad  \mathcal{\tilde Q}_{5|I}  = \int { \mathcal{\tilde J}_{5|I} }.\label{Charges}
\ee

The geometric way of understanding these charges is noting that
they descend from the eleven dimensional electric and magnetic
M-brane charges, hence the $\left(2,5\right)$ labels\footnote{This
is the reverse situation to that of \cite{Gutperle:2000ve}, where
the (dual) Euclidean theory was studied.}. M2-branes wrapping
special Lagrangian cycles of $\M$ generate $\mathcal{Q}_2$ while
the wrapping of M5-branes excite
$\left(\mathcal{Q}_5^I,\mathcal{\tilde Q}_{5|I}\right)$.

Finally, for completeness sake we also give $da$, where $a$ is the
universal axion dual to $A$. Since (\ref{Feomgeneral}) is
equivalent to $d^2 a =0$, we conclude that
\be\label{UniversalAxion}
    da = e^{ - 2\sigma } \star F - \left( {\zeta ^I d\tilde \zeta _I  - \tilde \zeta _I d\zeta ^I } \right),
\ee
where $a$ is governed by the field equation
\be\label{a field equation}
    d^{\dag} \left[ {e^{2\sigma } da + e^{2\sigma } \left( {\zeta ^I d\tilde \zeta _I  - \tilde \zeta _I d\zeta ^I } \right)} \right] =    0;
\ee
as a consequence of $dF=0$. Both terms involving $F$ in
(\ref{Fullaction}) could then be replaced by the single
expression\footnote{Alternatively, one may dualize the action by
introducing $a$ as a Lagrange multiplier and modifying the action
accordingly \cite{Gutperle:2000ve}.}
\be
    S_a  = \frac{1}{2}\int {e^{2\sigma } \left[ {da + \left( {\zeta^I d\tilde \zeta_I  - \tilde \zeta_I d\zeta^I } \right)} \right] \wedge \star\left[ {da + \left( {\zeta^I d\tilde \zeta_I  - \tilde \zeta_I d\zeta^I } \right)}
    \right]}.\label{a action}
\ee

The full supersymmetric action is invariant under the following SUSY variations. For the gravitini:
\begin{eqnarray}
    \delta_\epsilon \psi  ^A  &=& \tilde{\nabla}  \epsilon^A + \left[ {\mathcal{G}  } \right]_{\;\;B}^A \epsilon ^B  \nonumber\\
    \left[ {\mathcal{G}  } \right] &=& \left[ {\begin{array}{*{20}c}
    {\frac{1}{4}\left( {v   - \bar v   - Y  } \right)} & { - \bar
    u     }  \\
    {u     } & { - \frac{1}{4}\left( {v   - \bar v   - Y  } \right)}
    \\
    \end{array}} \right] \nonumber \\ \label{gravitinotrans}
\end{eqnarray}
where the indices $A$ and $B$ run over $(1,2)$, $\tilde \nabla$ is given
by
\be
    \tilde{\nabla}=dx^\mu\left( \partial _\mu   + \frac{1}{4}\omega _\mu^{\,\,\,\,\hat \mu\hat \nu} \Gamma _{\hat \mu\hat
    \nu}\right)
\ee
where the $\omega$'s are the usual spin connections, hated indices denote dimensions in a flat tangent space and the $\epsilon$'s are the SUSY parameters. The other quantities in (\ref{gravitinotrans}) are
\begin{eqnarray}
    u   &=&  e^{\frac{\sigma }{2}} \left( {M_I  {d    \zeta^I } + L^I  {d  \tilde \zeta _I } }    \right) \quad\quad\quad
    \bar u   = e^{\frac{\sigma }{2}} \left( {\bar M_I  {d \zeta ^I }  + \bar L^I  {d  \tilde \zeta _I }    } \right) \nonumber \\
    v   &=& \frac{1}{2} {d  \sigma }    + \frac{i}{2}e^{-\sigma}  \star F \quad\quad\quad\quad\quad
    \bar v   = \frac{1}{2} {d  \sigma }    - \frac{i}{2}e^{-\sigma}  \star F\label{eqns5}
\end{eqnarray}
and
\begin{equation}
    Y   = \frac{{\bar Z^I N_{IJ}  {d  Z^J }  -
    Z^I N_{IJ}  {d  \bar Z^J } }}{{\bar Z^I N_{IJ} Z^J
    }}
\end{equation}
which is proportional to the $U\left(1\right)$ K\"{a}hler connection defined by (\ref{U1connection2}).

Finally, the hyperini equations are:
\be
    \delta_\epsilon \xi _1^I  = e_{\;\;\mu} ^{1I} \Gamma ^\mu  \epsilon _1  - \bar e_{\;\;\mu}^{2I}
    \Gamma ^\mu  \epsilon _2,  \quad\quad\quad\quad
    \delta_\epsilon \xi _2^I  = e_{\;\;\mu} ^{2I} \Gamma ^\mu  \epsilon _1  + \bar e_{\;\;\mu}^{1I}
    \Gamma ^\mu  \epsilon _2, \label{hyperinotrans}
\ee
written in terms of the quantities:
\be
    e ^{1I}=e_{\;\;\mu} ^{1I}dx^\mu  = \left( {\begin{array}{*{20}c}
   {u  }  \\
   {E ^{\hat i} }  \\
    \end{array}} \right) \nonumber \\ , \quad\quad\quad
    e^{2I}=e_{\;\;\mu} ^{2I}dx^\mu  = \left(
    {\begin{array}{*{20}c}
   {v  }  \\
   {e ^{\hat i} }  \\
    \end{array}} \right)
\ee
\be
    E ^{\hat i}  =  e^{\frac{\sigma }{2}} e^{\hat ij} \left( {h_{jI}    {d  \zeta ^I }  + f_j^I  {d  \tilde \zeta _I }    } \right), \quad\quad\quad
    \bar E ^{\hat i}  =  e^{\frac{\sigma }{2}} e^{\hat i\bar j} \left( {h_{\bar    jI}      {d \zeta ^I }  + f_{\bar j}^I  {d  \tilde \zeta _I } } \right)
\ee
and the beins of the special K\"{a}hler metric:
\be
    e ^{\hat i}  = e_{\;\;j}^{\hat i}  {d  z^j } \quad\quad \quad
    \quad \bar e^{\hat i}  = e_{\;\;{\bar j}}^{\hat i}  {d  z^{\bar j} } \quad\quad\quad
    G_{i\bar j}  = e_{\;\;i}^{\hat k} e_{\;\;{\bar j}}^{\hat l} \delta _{\hat k\hat l}.
\ee

\section{The theory in symplectic form\label{mathematicasymplectica}}

In this section we arrive at our main objective: recasting the
action (\ref{Fullaction}) and its associated field and SUSY
equations into a manifestly symplectic form based on the language
defined in \S\ref{SKGandSp}. The reader should be convinced by now
that this is a straightforward matter and can be achieved by
direct examination of the equations involved. We give as much
detail as possible for the sake of future reference. Finally, we
show how a calculation based on the symplectic formulation may be
carried out by direct application to the results of
\cite{Emam:2005bh} and \cite{Gutperle:2000ve}.

\subsection{Reformulation}

The action (\ref{Fullaction}) is invariant under rotations in \textbf{\textit{Sp}},
so by inspection it is clear that $R$, $d\sigma$, $dz$ and $F$
are themselves symplectic invariants, whose explicit form will
depend on the specific ans\"{a}tze used. The axion fields
$\left(\zeta, \tilde\zeta\right)$, however, can be thought of as
components of an \textbf{\textit{Sp}} ``axions vector''. If we
define:
\be\label{XiasSp}
   \left| \Xi  \right\rangle  = \left( {\begin{array}{*{20}c}
   {\,\,\,\,\,\zeta ^I }  \\
   -{\tilde \zeta _I }  \\
    \end{array}} \right), \quad\quad\quad\quad\left| {d\Xi } \right\rangle  = \left( {\begin{array}{*{20}c}
   {\,\,\,\,\,d\zeta ^I }  \\
   -{d\tilde \zeta _I }  \\
    \end{array}} \right)
\ee
then clearly
\be
    \left\langle {{\Xi }}
 \mathrel{\left | {\vphantom {{\Xi } d\Xi }}
 \right. \kern-\nulldelimiterspace}
 {d\Xi } \right\rangle   = \zeta^I d\tilde \zeta_I  - \tilde \zeta_I
 d\zeta^I,
\ee
as well as:
\bea
    \lefteqn{
    \left\langle {\partial _\mu  \Xi } \right|\Lambda \left| {\partial ^\mu  \Xi }
    \right\rangle
    }\nonumber\\
    &=&    -\left( {\gamma_{IJ}  + \gamma ^{ KL} \theta _{IK}\theta _{JL} } \right) \left( {\partial _\mu  \zeta ^I } \right) \left( {\partial ^\mu  \zeta ^J } \right)  - \gamma ^{IJ}  \left( {\partial _\mu  \tilde \zeta _I } \right) \left( {\partial ^\mu  \tilde \zeta _J } \right)  - 2\gamma ^{ IK} \theta_{JK}  \left( {\partial _\mu  \zeta ^J } \right) \left( {\partial ^\mu  \tilde \zeta _I }
    \right),\nonumber\\
\eea
such that (\ref{X}) becomes
\bea
    X&=& {\left( {\gamma_{IJ}  + \gamma ^{ KL} \theta _{IK}\theta _{JL} } \right) {d\zeta^I } \wedge\star {d\zeta^J }  + \gamma ^{IJ}  {d\tilde \zeta_I } \wedge\star {d\tilde \zeta_J }  + 2\gamma ^{ IK} \theta_{JK}  {d\zeta^J } \wedge\star {d\tilde \zeta_I } }
   \nonumber\\&=&  - \left\langle {\partial _\mu  \Xi } \right|\Lambda \left| {\partial ^\mu  \Xi } \right\rangle \star   1.
\eea

Also note that we chose the minus sign in the definition
(\ref{XiasSp}) such that the resulting equations agree with the
form of the theory used in previous work, particularly
\cite{Emam:2005bh,Emam:2006sr,Gutperle:2000ve}. Replacing the
minus sign with a positive sign would result in the appearance of
minus signs in various locations in the action, field and SUSY
equations.

As a consequence of this language, the field expansion
(\ref{FExpanded}) could be rewritten
\bea
    \A &=& A + \sqrt 2  \left\langle {\Theta }
 \mathrel{\left | {\vphantom {\Theta  \Xi }}
 \right. \kern-\nulldelimiterspace}
 {\Xi } \right\rangle,\nonumber\\
    \F &=& d\A = F + \sqrt 2 \mathop {\left\langle {\Theta }
 \mathrel{\left | {\vphantom {\Theta  {d\Xi }}}
 \right. \kern-\nulldelimiterspace}
 {{d\Xi }} \right\rangle }\limits_{ \wedge \,\,\,\,}. \label{FExpandedSp}
\eea

The bosonic action in manifest symplectic covariance is hence:
\bea
    S_5  &=& \int\limits_5 {\left[ {R\star 1 - \frac{1}{2}d\sigma \wedge\star d\sigma  - G_{i\bar j} dz^i \wedge\star dz^{\bar j} } \right.}  \nonumber\\
    & &\left. {- F \wedge \left\langle {{\Xi }}
    \mathrel{\left | {\vphantom {{\Xi } d\Xi }} \right. \kern-\nulldelimiterspace} {d\Xi } \right\rangle  - \frac{1}{2}e^{ - 2\sigma } F \wedge \star F
    + e^\sigma   \left\langle {\partial _\mu  \Xi } \right|\Lambda \left| {\partial ^\mu  \Xi } \right\rangle\star 1}
    \right].
\eea

The equations of motion are now
\bea
    \left( {\Delta \sigma } \right)\star 1 + e^\sigma   \left\langle {\partial _\mu  \Xi } \right|\Lambda \left| {\partial ^\mu  \Xi } \right\rangle \star 1 + e^{ - 2\sigma } F \wedge \star F &=& 0\label{dilatoneomSp}\\
    \left( {\Delta z^i } \right)\star 1 + \Gamma _{jk}^i dz^j  \wedge \star dz^k  + \frac{1}{2}e^\sigma  G^{i\bar j}  {\partial _{\bar j} \left\langle {\partial
    _\mu \Xi}    \right|\Lambda \left| {\partial ^\mu  \Xi }   \right\rangle\star  1} &=& 0 \nonumber\\
    \left( {\Delta z^{\bar i} } \right)\star 1 + \Gamma _{\bar j\bar k}^{\bar i} dz^{\bar j}  \wedge \star dz^{\bar k}  + \frac{1}{2}e^\sigma  G^{\bar ij}  {\partial _j \left\langle {\partial _\mu
    \Xi}   \right|\Lambda \left| {\partial ^\mu  \Xi } \right\rangle\star 1}  &=& 0\label{zeomSp} \\
    d^{\dag} \left[ {e^{ - 2\sigma } F + \star\left\langle {{\Xi }} \mathrel{\left | {\vphantom {{\Xi } d\Xi }}
     \right. \kern-\nulldelimiterspace} {d\Xi } \right\rangle} \right] &=& 0\label{FeomgeneralSp}\\
    d^\dag\left[ {e^\sigma  \left| {\Lambda d\Xi } \right\rangle  +  \star F \left| {\Xi } \right\rangle } \right] &=&0.\label{xieomSp}
\eea

Note that, as is usual for Chern-Simons actions, the explicit
appearance of the gauge potential $\left| \Xi  \right\rangle $ in
(\ref{FeomgeneralSp}) and (\ref{xieomSp}) does not have an effect
on the physics since:
\bea
    d^{\dag} \star\left\langle {\Xi }
    \mathrel{\left | {\vphantom {\Xi  {d\Xi }}}
    \right. \kern-\nulldelimiterspace}
    {{d\Xi }} \right\rangle  & &\longrightarrow\quad d\left\langle {\Xi }
    \mathrel{\left | {\vphantom {\Xi  {d\Xi }}}
    \right. \kern-\nulldelimiterspace}
    {{d\Xi }} \right\rangle  = \mathop {\left\langle {{d\Xi }}
    \mathrel{\left | {\vphantom {{d\Xi } {d\Xi }}}
    \right. \kern-\nulldelimiterspace}
    {{d\Xi }} \right\rangle }\limits_ \wedge\nonumber\\
    d^{\dag} \star F\left| \Xi  \right\rangle  & &\longrightarrow\quad d\left[ {F\left| \Xi  \right\rangle } \right] = F\wedge \left| {d\Xi } \right\rangle,
\eea
where the Bianchi identities on $A$ and ${\left| \Xi \right\rangle
}$ were used. Now, if $\left| \Xi \right\rangle $ is taken to be
independent of the moduli, then we can write
\be
    \partial _j \left\langle {\partial _\mu  \Xi } \right|\Lambda \left| {\partial ^\mu  \Xi } \right\rangle  = \left\langle {\partial _\mu  \Xi } \right|\partial _j \Lambda \left| {\partial ^\mu  \Xi } \right\rangle.
\ee

Furthermore, since the exponents of both the $\M_C$ K\"{a}hler
potential $\K$ and the dilaton $\sigma$ are proportional to the
volume of the CY submanifold, then they can be taken to be
proportional to each other, \emph{i.e.} following
\cite{Sabra:1997dh}:
\be
    \sigma = c \K,
\ee
where $c$ is some arbitrary constant. The Noether currents and
charges become
\bea
    \mathcal{J}_2  &=& e^{ - 2\sigma } F + \star\left\langle {{\Xi }} \mathrel{\left | {\vphantom {{\Xi } d\Xi }}
     \right. \kern-\nulldelimiterspace} {d\Xi } \right\rangle\nonumber\\
     \left| {\mathcal{J}_5 } \right\rangle &=&e^\sigma  \left| {\Lambda d\Xi } \right\rangle  +  \star F \left| {\Xi } \right\rangle\nonumber\\
    \mathcal{Q}_2  &=& \int {\mathcal{J}_2 },\quad \quad \quad
    \left| {\mathcal{Q}_5 } \right\rangle  = \int {\left| {\mathcal{J}_5 } \right\rangle }.
    \label{CurrentsChargesSp}
\eea

The equations of the universal axion (\ref{UniversalAxion}),
(\ref{a field equation}) and (\ref{a action}) are now
\be
    da = e^{ - 2\sigma } \star F - \left\langle {\Xi } \mathrel{\left | {\vphantom {\Xi  {d\Xi }}} \right. \kern-\nulldelimiterspace} {{d\Xi }}
    \right\rangle,
\ee
\be
    d^{\dag} \left[ {e^{2\sigma } da + e^{2\sigma } \left\langle {\Xi } \mathrel{\left | {\vphantom {\Xi  {d\Xi }}} \right. \kern-\nulldelimiterspace} {{d\Xi }}
    \right\rangle} \right] =    0\quad\quad {\rm and}
\ee
\be
    S_a  = \frac{1}{2}\int {e^{2\sigma } \left[ {da + \left\langle {\Xi } \mathrel{\left | {\vphantom {\Xi  {d\Xi }}} \right. \kern-\nulldelimiterspace} {{d\Xi }}
    \right\rangle} \right] \wedge \star\left[ {da + \left\langle {\Xi } \mathrel{\left | {\vphantom {\Xi  {d\Xi }}} \right. \kern-\nulldelimiterspace} {{d\Xi }}
    \right\rangle}    \right]}.
\ee

Next, we look at the SUSY variations. The gravitini equations can
be explicitly written as follows:
\bea
 \delta _\epsilon  \psi ^1  &=& \tilde{\nabla} \epsilon _1  + \frac{1}{4}\left( {ie^{ - \sigma } \star F - Y} \right)\epsilon _1  - e^{\frac{\sigma }{2}} \left\langle {{\bar V}}
 \mathrel{\left | {\vphantom {{\bar V} {d\Xi }}} \right. \kern-\nulldelimiterspace} {{d\Xi }} \right\rangle\epsilon _2  \\
 \delta _\epsilon  \psi ^2  &=& \tilde{\nabla} \epsilon _2  - \frac{1}{4}\left( {ie^{ - \sigma } \star F - Y} \right)\epsilon _2  + e^{\frac{\sigma }{2}} \left\langle {V}
 \mathrel{\left | {\vphantom {V {d\Xi }}} \right. \kern-\nulldelimiterspace} {{d\Xi }} \right\rangle \epsilon _1,\label{SUSYSpGravitini}
\eea
while the hyperini variations are
\bea
     \delta _\epsilon  \xi _1^0  &=& e^{\frac{\sigma }{2}} \left\langle {V}
    \mathrel{\left | {\vphantom {V {\partial _\mu  \Xi }}} \right. \kern-\nulldelimiterspace} {{\partial _\mu  \Xi }} \right\rangle  \Gamma ^\mu  \epsilon _1  - \left[ {\frac{1}{2}\left( {\partial _\mu  \sigma } \right) - \frac{i}{2}e^{ - \sigma } \left( {\star F} \right)_\mu  } \right]\Gamma ^\mu  \epsilon _2  \nonumber\\
     \delta _\epsilon  \xi _2^0  &=& e^{\frac{\sigma }{2}} \left\langle {{\bar V}}
    \mathrel{\left | {\vphantom {{\bar V} {\partial _\mu  \Xi }}} \right. \kern-\nulldelimiterspace} {{\partial _\mu  \Xi }} \right\rangle \Gamma ^\mu  \epsilon _2  + \left[ {\frac{1}{2}\left( {\partial _\mu  \sigma } \right) + \frac{i}{2}e^{ - \sigma } \left( {\star F} \right)_\mu  } \right]\Gamma ^\mu  \epsilon
     _1 \label{SUSYSpHyperiniFirst}\\
     \delta _\epsilon  \xi _1^{\hat i}  &=& e^{\frac{\sigma }{2}} e^{\hat ij} \left\langle {{U_j }}
    \mathrel{\left | {\vphantom {{U_j } {\partial _\mu  \Xi }}} \right. \kern-\nulldelimiterspace} {{\partial _\mu  \Xi }} \right\rangle \Gamma ^\mu  \epsilon _1  - e_{\,\,\,\bar j}^{\hat i} \left( {\partial _\mu  z^{\bar j} } \right)\Gamma ^\mu  \epsilon _2  \nonumber\\
     \delta _\epsilon  \xi _2^{\hat i}  &=& e^{\frac{\sigma }{2}} e^{\hat i\bar j} \left\langle {{U_{\bar j} }}
    \mathrel{\left | {\vphantom {{U_{\bar j} } {\partial _\mu  \Xi }}} \right. \kern-\nulldelimiterspace} {{\partial _\mu  \Xi }} \right\rangle \Gamma ^\mu  \epsilon _2  + e_{\,\,\,j}^{\hat i} \left( {\partial _\mu  z^j } \right)\Gamma ^\mu  \epsilon
     _1.\label{SUSYSpHyperini}
\eea

For easy reference, we also compute:
\bea
    dG_{i\bar j}  &=& G_{k\bar j} \Gamma _{ri}^k dz^r  + G_{i\bar k} \Gamma _{\bar r\bar j}^{\bar k} dz^{\bar r}  \nonumber\\
    dG^{i\bar j}  &=&  - G^{p\bar j} \Gamma _{rp}^i dz^r  - G^{i\bar p} \Gamma _{\bar r\bar p}^{\bar j} dz^{\bar r}  \nonumber\\
    \left| {dV} \right\rangle  &=& dz^i \left| {U_i } \right\rangle  - i\mathcal{P}\left| V \right\rangle \nonumber \\
    \left| {d\bar V} \right\rangle  &=& dz^{\bar i} \left| {U_{\bar i} } \right\rangle  + i\mathcal{P}\left| {\bar V} \right\rangle \nonumber \\
    \left| {dU_i } \right\rangle  &=& G_{i\bar j} dz^{\bar j} \left| V \right\rangle  + \Gamma _{ik}^r dz^k \left| {U_r } \right\rangle+G^{j\bar l} C_{ijk} dz^k \left| {U_{\bar l} } \right\rangle - i\mathcal{P}\left| {U_i } \right\rangle \nonumber \\
    \left| {dU_{\bar i} } \right\rangle  &=& G_{j\bar i} dz^j \left| {\bar V} \right\rangle + \Gamma _{\bar i\bar k}^{\bar r} dz^{\bar k} \left| {U_{\bar r} } \right\rangle + G^{l\bar j} C_{\bar i\bar j\bar k} dz^{\bar k} \left| {U_l } \right\rangle + i\mathcal{P}\left| {U_{\bar i} } \right\rangle \nonumber \\
    d{\bf \Lambda } &=& \left( {\partial _i {\bf \Lambda }} \right)dz^i  + \left( {\partial _{\bar i} {\bf \Lambda }} \right)dz^{\bar i},\label{SpacetimeVariations}
\eea
where $\mathcal{P}$ is the $U\left(1\right)$ connection defined by
(\ref{U1connection2}) and $\left( {\partial _i {\bf \Lambda }} , {\partial _{\bar i}
{\bf \Lambda }}\right)$ are given by (\ref{CovariantDerofLambda}).

\subsection{Examples}\label{Application}

The analysis of solution ans\"{a}tze representing hypermultiplet
fields should now reduce to the problem of constructing and
manipulating symplectic quantities. Using the language developed
in this paper, we now demonstrate how this can be done by applying
the symplectic method to two known results.

In \cite{Emam:2005bh,Emam:2006sr} we studied the
dimensional reduction of M5-branes wrapping special Lagrangian
cycles of a Calabi-Yau 3-fold and showed explicitly that it led to
Bogomol'nyi-Prasad-Sommerfield (BPS) 2-branes coupled to the five dimensional $\N=2$
hypermultiplets with constant universal axion ($F=da=0$). The case
with nontrivial complex structure moduli led to constraint
equations on the solution that turned out to be of the attractor
type. We will not reproduce the entire calculation here, but
rather only show enough to demonstrate how the symplectic method
greatly reduces the effort involved.

The $D=5$ spacetime metric due to the presence of the 2-brane was
found to be of the form
\be
    ds^2  = \left( { - dt^2  + dx_1^2  + dx_2^2 } \right) + e^{ - 2\sigma } \left( {dx_3^2  + dx_4^2 }
    \right),
\ee
where $\left(x^1,x^2\right)$ define the spatial directions tangent
to the brane and $\left(x^3,x^4\right)$ define those transverse to
it. The constraint equations on the dilaton and moduli are
\bea
 d\left( {e^{ - \frac{\sigma }{2}} } \right) &=& \left\langle {{d\mathcal{H}}}
 \mathrel{\left | {\vphantom {{d\mathcal{H}} V}}
 \right. \kern-\nulldelimiterspace}
 {V} \right\rangle  = \left\langle {{d\mathcal{H}}}
 \mathrel{\left | {\vphantom {{d\mathcal{H}} {\bar V}}}
 \right. \kern-\nulldelimiterspace}
 {{\bar V}} \right\rangle  \nonumber \\
 dz^i  &=& - e^{\frac{\sigma }{2}}  G^{i\bar j} \left\langle {{d\mathcal{H}}}
 \mathrel{\left | {\vphantom {{d\mathcal{H}} {U_{\bar j} }}}
 \right. \kern-\nulldelimiterspace}
 {{U_{\bar j} }} \right\rangle \nonumber \\
 dz^{\bar i}  &=& - e^{\frac{\sigma }{2}} G^{\bar ij} \left\langle {{d\mathcal{H}}}
 \mathrel{\left | {\vphantom {{d\mathcal{H}} {U_j }}}
 \right. \kern-\nulldelimiterspace}
 {{U_j }} \right\rangle\label{D5Solution}
\eea
where
\be
    \left| \mathcal{H} \right\rangle  = \left( {\begin{array}{*{20}c}
   {H^I }  \\
   {\tilde H_I }  \\
    \end{array}} \right)\label{Harmonicfunctionvector}
\ee
is taken to be dependent only on the $\left(x^3,x^4\right)$
coordinates, such that the moduli dependence is carried
exclusively by $\left| V \right\rangle$ and  $\left| U
\right\rangle $. The field equations are straightforwardly
satisfied if $\left| \mathcal{H} \right\rangle$ is
taken to be radial and harmonic in the transverse plane, \emph{i.e.}
\be
    \left| {\Delta \mathcal{H}} \right\rangle  = 0,
\ee
which is generally solved by
\be
    \left| \mathcal{H} \right\rangle  = \left| \lambda  \right\rangle  + \ln r\left| \varpi
    \right\rangle,\label{Harmonicfunctionssolution}
\ee
where $r$ is the radial coordinate in the $\left(x^3,x^4\right)$
plane, $\left| \lambda \right\rangle$ is an arbitrary constant and
\be
    \left| \varpi\right\rangle  = \left( {\begin{array}{*{20}c}
   {q^I }  \\
   {\tilde q_I }  \\
    \end{array}} \right),\label{Chargesvector}
\ee
defines constant ``electric'' and ``magnetic'' charges excited by the
wrapping of the M5-brane over each homology cycle on the
submanifold $\M$. It follows then that
\bea
 \left| {d\mathcal{H}} \right\rangle  &=& \frac{{dr}}{r}\left| \varpi \right\rangle  \quad{\rm and}\quad
 \left| {\star d\mathcal{H}} \right\rangle  = d\varphi\left| \varpi
 \right\rangle,
\eea
where $\varphi$ is the angular coordinate in the
$\left(x^3,x^4\right)$ plane. We take the axions vector to be of
the simple form
\be
    \left| {d\Xi } \right\rangle  =\pm\left| {\star d\mathcal{H} } \right\rangle=  \pm d\varphi \left| \varpi
    \right\rangle.\label{Axion Ansatz}
\ee

The dilaton equation (\ref{dilatoneomSp}) is now:
\be
    \left( {\Delta \sigma } \right)\star 1 + e^\sigma  \left\langle {d\Xi } \right|\mathop \Lambda \limits_ \wedge  \left| {\star d\Xi }
    \right\rangle=0.\label{DilatonSolving}
\ee

The first term of (\ref{DilatonSolving}) gives
\be
    \left( {\Delta \sigma } \right)\star 1 =  - 2e^\sigma  \left\langle {{\star d\mathcal{H}}}
 \mathrel{\left | {\vphantom {{\star d\mathcal{H}} V}}
 \right. \kern-\nulldelimiterspace}
 {V} \right\rangle  \wedge \left\langle {{\bar V}}
 \mathrel{\left | {\vphantom {{\bar V} {d\mathcal{H}}}}
 \right. \kern-\nulldelimiterspace}
 {{d\mathcal{H}}} \right\rangle  - 2e^\sigma  G^{i\bar j} \left\langle {{\star d\mathcal{H}}}
 \mathrel{\left | {\vphantom {{\star d\mathcal{H}} {U_{\bar j} }}}
 \right. \kern-\nulldelimiterspace}
 {{U_{\bar j} }} \right\rangle  \wedge \left\langle {{U_i }}
 \mathrel{\left | {\vphantom {{U_i } {d\mathcal{H}}}}
 \right. \kern-\nulldelimiterspace}
 {{d\mathcal{H}}} \right\rangle.
\ee

Now, with the knowledge that
\be
    \mathop {\left\langle {{\star d\mathcal{H}}}
    \mathrel{\left | {\vphantom {{\star d\mathcal{H}} {d\mathcal{H}}}}
    \right. \kern-\nulldelimiterspace}
    {{d\mathcal{H}}} \right\rangle }\limits_ \wedge   \propto \left\langle
    {\varpi }
    \mathrel{\left | {\vphantom {\varpi  \varpi }}
    \right. \kern-\nulldelimiterspace}
    {\varpi } \right\rangle  = 0,
\ee
as well as
\be
    \left\langle {d\Xi } \right|\mathop \Lambda \limits_ \wedge  \left| {\star d\Xi } \right\rangle  = 2\left\langle {{\star d\mathcal{H}}}
    \mathrel{\left | {\vphantom {{\star d\mathcal{H}} V}}
    \right. \kern-\nulldelimiterspace}
    {V} \right\rangle  \wedge \left\langle {{\bar V}}
    \mathrel{\left | {\vphantom {{\bar V} {d\mathcal{H}}}}
    \right. \kern-\nulldelimiterspace}
    {{d\mathcal{H}}} \right\rangle  + 2G^{i\bar j} \left\langle {{\star d\mathcal{H}}}
    \mathrel{\left | {\vphantom {{\star d\mathcal{H}} {U_{\bar j} }}}
    \right. \kern-\nulldelimiterspace}
    {{U_{\bar j} }} \right\rangle  \wedge \left\langle {{U_i }}
    \mathrel{\left | {\vphantom {{U_i } {d\mathcal{H}}}}
    \right. \kern-\nulldelimiterspace}
    {{d\mathcal{H}}} \right\rangle,
\ee
it is clear that the second term of (\ref{DilatonSolving}) exactly
cancels the first.

The moduli equations involve a slightly longer calculation. The
first term of (\ref{zeomSp}) gives
\bea
    \left( {\Delta z^i } \right)\star 1 &=& e^\sigma  G^{i\bar j} G^{l\bar m} G^{k\bar n} C_{\bar j\bar m\bar n} \left\langle {{\star d\mathcal{H}}}
    \mathrel{\left | {\vphantom {{\star d\mathcal{H}} {U_l }}}
    \right. \kern-\nulldelimiterspace}
    {{U_l }} \right\rangle  \wedge \left\langle {{d\mathcal{H}}}
    \mathrel{\left | {\vphantom {{d\mathcal{H}} {U_k }}}
    \right. \kern-\nulldelimiterspace}
    {{U_k }} \right\rangle
    + e^\sigma  G^{i\bar j} \left\langle {{\star d\mathcal{H}}}
    \mathrel{\left | {\vphantom {{\star d\mathcal{H}} {\bar V}}}
    \right. \kern-\nulldelimiterspace}
    {{\bar V}} \right\rangle  \wedge \left\langle {{d\mathcal{H}}}
    \mathrel{\left | {\vphantom {{d\mathcal{H}} {U_{\bar j} }}}
    \right. \kern-\nulldelimiterspace}
    {{U_{\bar j} }} \right\rangle  \nonumber\\
     &+& e^\sigma  G^{i\bar j}
    \left\langle {{d\mathcal{H}}}
    \mathrel{\left | {\vphantom {{d\mathcal{H}} V}}
    \right. \kern-\nulldelimiterspace}
    {V} \right\rangle  \wedge \left\langle {{\star d\mathcal{H}}}
    \mathrel{\left | {\vphantom {{\star d\mathcal{H}} {U_{\bar j} }}}
    \right. \kern-\nulldelimiterspace}
    {{U_{\bar j} }} \right\rangle
     - e^\sigma  G^{p\bar j} G^{r\bar k} \Gamma _{rp}^i \left\langle {{d\mathcal{H}}}
    \mathrel{\left | {\vphantom {{d\mathcal{H}} {U_{\bar k} }}}
    \right. \kern-\nulldelimiterspace}
    {{U_{\bar k} }} \right\rangle  \wedge \left\langle {{\star d\mathcal{H}}}
    \mathrel{\left | {\vphantom {{\star d\mathcal{H}} {U_{\bar j} }}}
    \right. \kern-\nulldelimiterspace}
    {{U_{\bar j} }} \right\rangle. \label{modulisolving1}
\eea

The second term is
\be
    \Gamma _{rp}^i dz^r  \wedge \star dz^p  = e^\sigma  G^{p\bar j} G^{r\bar k} \Gamma _{rp}^i \left\langle {{d\mathcal{H}}}
    \mathrel{\left | {\vphantom {{d\mathcal{H}} {U_{\bar k} }}}
    \right. \kern-\nulldelimiterspace}
    {{U_{\bar k} }} \right\rangle  \wedge \left\langle {{\star d\mathcal{H}}}
    \mathrel{\left | {\vphantom {{\star d\mathcal{H}} {U_{\bar j} }}}
    \right. \kern-\nulldelimiterspace}
    {{U_{\bar j} }} \right\rangle,
\ee
which cancels the last term of (\ref{modulisolving1}). Using
(\ref{CovariantDerofLambda}), the last term of the moduli equation
becomes
\bea
    & & \frac{1}{2}e^\sigma  G^{i\bar j} \left\langle {d\Xi }
    \right|\mathop {\partial _{\bar j} \Lambda }\limits_ \wedge \left|
    {\star d\Xi } \right\rangle  =  - e^\sigma  G^{i\bar j} G^{l\bar m}
    G^{k\bar n} C_{\bar j\bar m\bar n} \left\langle {{\star d\mathcal{H}}}
    \mathrel{\left | {\vphantom {{\star d\mathcal{H}} {U_l }}}
    \right. \kern-\nulldelimiterspace}
    {{U_l }} \right\rangle  \wedge \left\langle {{d\mathcal{H}}}
    \mathrel{\left | {\vphantom {{d\mathcal{H}} {U_k }}}
    \right. \kern-\nulldelimiterspace}
    {{U_k }} \right\rangle  \nonumber\\
    & & - e^\sigma  G^{i\bar j} \left\langle {{\star d\mathcal{H}}}
    \mathrel{\left | {\vphantom {{\star d\mathcal{H}} {\bar V}}}
    \right. \kern-\nulldelimiterspace}
    {{\bar V}} \right\rangle  \wedge \left\langle {{d\mathcal{H}}}
    \mathrel{\left | {\vphantom {{d\mathcal{H}} {U_{\bar j} }}}
    \right. \kern-\nulldelimiterspace}
    {{U_{\bar j} }} \right\rangle
     - e^\sigma  G^{i\bar j}
    \left\langle {{d\mathcal{H}}}
    \mathrel{\left | {\vphantom {{d\mathcal{H}} V}}
    \right. \kern-\nulldelimiterspace}
    {V} \right\rangle  \wedge \left\langle {{\star d\mathcal{H}}}
    \mathrel{\left | {\vphantom {{\star d\mathcal{H}} {U_{\bar j} }}}
    \right. \kern-\nulldelimiterspace}
    {{U_{\bar j} }} \right\rangle,
\eea
exactly canceling the remaining terms of (\ref{modulisolving1}).

The second example we wish to consider is that of \cite{Gutperle:2000ve}. The result discussed therein was that of instanton couplings to the hypermultiplets. Instantons are of course Euclidean solutions of the theory and may be thought of as being magnetically dual to the 2-branes discussed above (in $D=5$). In order to consider this result, we analytically continue the action of the theory from a Minkowski background to a Euclidean metric. This is achieved by an ordinary Wick rotation which has the effect of changing $\left| \Xi  \right\rangle  \to i\left| \Xi  \right\rangle $ in the field and SUSY equations. Furthermore, the vector $\left| \mathcal{H} \right\rangle $ satisfying the harmonic condition in Euclidean $D=5$ space now becomes
\be
    \left| \mathcal{H} \right\rangle  = \left| \lambda  \right\rangle  + \frac{1}{{3r^3 }}\left| \varpi  \right\rangle,
\ee
instead of (\ref{Harmonicfunctionssolution}), with (\ref{Chargesvector}) still valid. Note that the coordinate $r$ is now radial in all the five flat dimensions. Hence
\be
   \left| {d\mathcal{H}} \right\rangle  =  - \frac{{dr}}{{r^4 }}\left| \varpi  \right\rangle.
\ee

Rewriting the constraint equations on the dilaton and moduli in our language we get:
\bea
 d\left( {e^{ \frac{\sigma }{2}} } \right) &=& -\left\langle {{d\mathcal{H}}}
 \mathrel{\left | {\vphantom {{d\mathcal{H}} V}}
 \right. \kern-\nulldelimiterspace}
 {V} \right\rangle  = -\left\langle {{d\mathcal{H}}}
 \mathrel{\left | {\vphantom {{d\mathcal{H}} {\bar V}}}
 \right. \kern-\nulldelimiterspace}
 {{\bar V}} \right\rangle  \nonumber \\
 dz^i  &=& e^{-\frac{\sigma }{2}}  G^{i\bar j} \left\langle {{d\mathcal{H}}}
 \mathrel{\left | {\vphantom {{d\mathcal{H}} {U_{\bar j} }}}
 \right. \kern-\nulldelimiterspace}
 {{U_{\bar j} }} \right\rangle \nonumber \\
 dz^{\bar i}  &=& e^{-\frac{\sigma }{2}} G^{\bar ij} \left\langle {{d\mathcal{H}}}
 \mathrel{\left | {\vphantom {{d\mathcal{H}} {U_j }}}
 \right. \kern-\nulldelimiterspace}
 {{U_j }} \right\rangle\label{D5SolutionInstanton}
\eea

while the axions can be written in the form
\be
    \left| {d\Xi } \right\rangle  =  - ie^{ - \sigma } \left| {\Lambda d\mathcal{H}} \right\rangle.
\ee

Now the dilaton and moduli equations can be shown to be satisfied in a very similar manner as that of the first example and the $\left| \Xi  \right\rangle$ field equation reduces to the harmonic condition on $\left| {\mathcal{H}} \right\rangle$:
\be
    d^{\dag} \left[ {e^\sigma  \left| {\Lambda d\Xi } \right\rangle } \right] =  - id^{\dag}  { \left| {\Lambda \Lambda d\mathcal{H}} \right\rangle }  = i\left| {\Delta \mathcal{H}} \right\rangle  = 0,
\ee
where the fact that $\Lambda ^{ - 1}  =  - \Lambda $ was used. The hyperini variations (\ref{SUSYSpHyperiniFirst}) and (\ref{SUSYSpHyperini}) vanish for $\epsilon_1 = \pm\epsilon_2$ as follows:
\bea
 \delta _\epsilon  \xi _1^0  &=&  - ie^{ - \frac{\sigma }{2}} \left\langle V \right|\Lambda \left| {d\mathcal{H}} \right\rangle  + e^{ - \frac{\sigma }{2}} \left\langle {V}
 \mathrel{\left | {\vphantom {V {d\mathcal{H}}}}
 \right. \kern-\nulldelimiterspace}
 {{d\mathcal{H}}} \right\rangle  \nonumber\\
  &=&  - i2e^{ - \frac{\sigma }{2}} \left\langle {V}
 \mathrel{\left | {\vphantom {V V}}
 \right. \kern-\nulldelimiterspace}
 {V} \right\rangle \left\langle {{\bar V}}
 \mathrel{\left | {\vphantom {{\bar V} {d\mathcal{H}}}}
 \right. \kern-\nulldelimiterspace}
 {{d\mathcal{H}}} \right\rangle  - i2e^{ - \frac{\sigma }{2}} G^{i\bar j} \left\langle {V}
 \mathrel{\left | {\vphantom {V {U_{\bar j} }}}
 \right. \kern-\nulldelimiterspace}
 {{U_{\bar j} }} \right\rangle \left\langle {{U_i }}
 \mathrel{\left | {\vphantom {{U_i } {d\mathcal{H}}}}
 \right. \kern-\nulldelimiterspace}
 {{d\mathcal{H}}} \right\rangle  \nonumber\\
  & & - e^{ - \frac{\sigma }{2}} \left\langle {V}
 \mathrel{\left | {\vphantom {V {d\mathcal{H}}}}
 \right. \kern-\nulldelimiterspace}
 {{d\mathcal{H}}} \right\rangle  + e^{ - \frac{\sigma }{2}} \left\langle {V}
 \mathrel{\left | {\vphantom {V {d\mathcal{H}}}}
 \right. \kern-\nulldelimiterspace}
 {{d\mathcal{H}}} \right\rangle  = 0,
\eea
where (\ref{Normality}) was used. Also
\bea
 \delta _\epsilon  \xi _1^{\hat i}  &=&  - ie^{ - \frac{\sigma }{2}} e^{\hat ij} \left\langle {U_j } \right|\Lambda \left| {d\mathcal{H}} \right\rangle  - e^{ - \frac{\sigma }{2}} e_{\,\,\bar k}^{\hat i} G^{\bar kj} \left\langle {{U_j }}
 \mathrel{\left | {\vphantom {{U_j } {d\mathcal{H}}}}
 \right. \kern-\nulldelimiterspace}
 {{d\mathcal{H}}} \right\rangle  \nonumber\\
  &=&  - i2e^{ - \frac{\sigma }{2}} e^{\hat ij} \left\langle {{U_j }}
 \mathrel{\left | {\vphantom {{U_j } V}}
 \right. \kern-\nulldelimiterspace}
 {V} \right\rangle \left\langle {{\bar V}}
 \mathrel{\left | {\vphantom {{\bar V} {d\mathcal{H}}}}
 \right. \kern-\nulldelimiterspace}
 {{d\mathcal{H}}} \right\rangle  - i2e^{ - \frac{\sigma }{2}} e^{\hat ij} G^{m\bar n} \left\langle {{U_j }}
 \mathrel{\left | {\vphantom {{U_j } {U_{\bar n} }}}
 \right. \kern-\nulldelimiterspace}
 {{U_{\bar n} }} \right\rangle \left\langle {{U_m }}
 \mathrel{\left | {\vphantom {{U_m } {d\mathcal{H}}}}
 \right. \kern-\nulldelimiterspace}
 {{d\mathcal{H}}} \right\rangle  \nonumber\\
  & &- e^{ - \frac{\sigma }{2}} e^{\hat ij} \left\langle {{U_j }}
 \mathrel{\left | {\vphantom {{U_j } {d\mathcal{H}}}}
 \right. \kern-\nulldelimiterspace}
 {{d\mathcal{H}}} \right\rangle  - e^{ - \frac{\sigma }{2}} e^{\hat ij} \left\langle {{U_j }}
 \mathrel{\left | {\vphantom {{U_j } {d\mathcal{H}}}}
 \right. \kern-\nulldelimiterspace}
 {{d\mathcal{H}}} \right\rangle  \nonumber\\
  &=& 2e^{ - \frac{\sigma }{2}} e^{\hat ij} \left\langle {{U_j }}
 \mathrel{\left | {\vphantom {{U_j } {d\mathcal{H}}}}
 \right. \kern-\nulldelimiterspace}
 {{d\mathcal{H}}} \right\rangle  - 2e^{ - \frac{\sigma }{2}} e^{\hat ij} \left\langle {{U_j }}
 \mathrel{\left | {\vphantom {{U_j } {d\mathcal{H}}}}
 \right. \kern-\nulldelimiterspace}
 {{d\mathcal{H}}} \right\rangle  = 0,
\eea
where (\ref{KmetricasSpproduct}) was used. Similarly $\delta _\epsilon  \xi _2^0 = 0$ and $\delta _\epsilon  \xi _2^{\hat i}=0$ are satisfied.

This is as far as we will go in demonstrating the use of the
symplectic method in analyzing the hypermultiplets. We note that the
calculations shown here are considerably shorter than their counterparts
performed without using the symplectic language. In fact, the original details
would indeed be too long to reasonably reproduce in print.

\section{Conclusion}

In this work, we took a close look at the geometries responsible
for the behavior of the hypermultiplet fields of five dimensional
$\N=2$ supergravity with particular emphasis on the symplectic
structure arising from the underlying topology of the Calabi-Yau
subspace. We proposed the use of the mathematics of symplectic
vector spaces to recast the theory in explicit symplectic
covariance. We argued that this greatly simplifies the effort
involved in analyzing the hypermultiplet fields, with or without
gravitational coupling, and demonstrated this by partially
applying it to two known results.

The five dimensional hypermultiplets sector is hardly the only one
exhibiting symplectic symmetry. In fact, the structures reviewed
here are almost always discussed in the literature in the
context of the four dimensional vector multiplets where very
similar analytical difficulties arise. In fact, it is because the
special K\"{a}hler geometry of the $D=4$ vector multiplets is so
well researched that it became possible to apply similar
techniques to the (c-mapped) $D=5$ hypermultiplets. It is then
natural to attempt to extend the symplectic formulation to the
$D=4$ theory as well as to any other theory, supersymmetric or
not, exhibiting hidden or explicit \textbf{\textit{Sp}}
covariance. One hopes that this will help simplify tedious
calculations as well as contribute to further understanding the
behavior of such theories.

Finally, an immediate application of the symplectic formulation to
analyzing solution ans\"{a}tze for various possible situations
seems to be the next natural thing to do. For example, an analysis
of branes coupled to the full set of hypermultiplet fields can now
be greatly simplified, even if one is interested in a general
understanding, rather than a detailed solution. Further
classification of such solutions becomes a more manageable task.
In the future, we plan to explore at least some of the above
possibilities.

\end{document}